# A Study of Long-term Energy-mix Optimization Model: A Case Study in Japan

Shintaro Negishi*

**Abstract** – There is a strong need to reduce greenhouse gas emissions to deal with climate change. In the power sector, changing the power generation method in the medium and long term is needed to reduce greenhouse gas emissions. This paper proposes a long-term energy-mix optimization model to obtain the process of carbon neutrality in the power system. The proposed model models power supply and demand at an hourly granularity and determines the generation capacity that minimizes the long-term energy supply cost. Compared with the models proposed in previous studies, the proposed model can determine the installed capacity to maintain the balance of power supply and demand by adding the capacity of regulation reserve required by fluctuations in the output of variable renewable energy as a constraint condition. A Japan energy mix calculation is reported as a case study of the proposed model. This model can clarify the roadmap to achieving each country's emission reduction target and support the government's decision-making.

## 1. Introduction

Countries worldwide are required to shift to a carbon-neutral energy supply system as a measure against climate change. Japan has declared itself carbon neutral by 2050 and has set a goal of reducing greenhouse gas emissions to 37% reduction by 2030 from 2013. To achieve this goal, decarbonizing the electricity sector is essential [1]. Although various industry associations have published roadmaps for achieving this goal (e.g., Japan Wind Power Association [2]), a quantitative roadmap from the overall optimization perspective is needed to ensure fair decision-making across the country.

The rapid spread of variable renewable energy (VRE) resources such as solar photovoltaic (PV) and wind power (WP) was motivated by a renewable energy feed-in tariff that went into effect in 2012, prior to the greenhouse gas reduction targets. It has been pointed out that issues such as the lack of various measures are considered to address these issues, including the introduction of grid storage batteries and demand control of consumer-side equipment. Economic calculations that include supply and demand control are necessary to evaluate the cost-effectiveness of these measures. Methods using the optimal power resource configuration model [3] and unit commitment problem (UC) [4] have been proposed to perform these economic calculations.

To clarify the path to achieving a carbon neutral state over several decades, developing a model is needed that can perform the operational analysis of supply and demand over a period of more than one year and optimize the capacity of various power sources and storage facilities.

Several optimization models have been proposed to determine the energy mix on a multi-year scale. Komiyama and Fujii [3] proposed a model that considers Japan as one area and determines the optimal energy resource capacities. H.C. Gils [4] proposed the "REMix" model, an optimal capacity model for electric power and thermal energy equipment. The "REMix" model is modeled by GAMS, an energy analysis software. The "REMix" model is also used to analyze energy supply and demand over a wide area in the EU [5]. In the optimization model for non-power systems, a capacity planning model for energy supply in small areas has been proposed [6].

These previous studies did not include regulation reserves to deal with fluctuations in the output of VRE. Regulation reserves are the "extra power" to maintain power supply and demand by using controllable thermal and hydroelectric power generation to cover the output of VRE that increases or decreases in seconds to minutes, depending on weather conditions.

Therefore, this paper proposes a long-term energy mix optimization model that includes regulation reserve capacity. The optimization results of Japan's energy mix are shown as a case study.

The contributions of this paper can be summarized as follows:

* Dept. of Electrical, Electronics and Information Engineering, Kanagawa University, Japan (negishi@kanagawa-u.ac.jp)

- **Energy-mix optimization model including regulation reserve capability:** The optimization model proposed in this paper ensures regulation reserve capability. Therefore, it can respond to frequency fluctuations and calculate the generation capacity to maintain a stable power supply and demand.
- **Possible to create a specific implementation plan to achieve carbon neutral state:** By using the proposed model, a year-by-year power supply transition plan toward the target year of carbon neutral state can be created. This contribution can identify energy technologies on which the government should focus its efforts.

The remainder of this paper is organized as follows. Section 2 describes the problem covered in this paper. Specifically, the environment assumed by the proposed model and the inputs and outputs of the model are described. Section 3 formulates the optimization problem for the long-term energy-mix optimization model. Section 4 describes a case study of a Japanese power system using the proposed model. This case study confirms the difference by including the regulation reserves. Section 5 discusses the results obtained in this paper.

## 2. Problem Description

Fig. 1 shows the model inputs/outputs and internal processing. Daily load curve, VRE installed capacity scenario, fuel price scenario, generator performance data, time-series data on VRE generation output, and hydro and geothermal/biomass generation output are the model input data. As a result of the optimization, the operating cost, the operation pattern of each generator, the capacity of the regulating power to be supplied, and the marginal hourly fuel cost are obtained.

## 3. Long-term Energy-mix Optimization Model

The objective function and constraints of the long-term energy mix optimization model are shown in (1)-(38). The proposed model is formulated as a linear programming problem. Table 1 lists the nomenclatures used in the model.

$$\min \sum_{y=1}^{yNum} \frac{1}{(1+r)^y} AC_y \quad (1)$$

subject to

$$AC_y = \sum_{i \in \mathcal{I}} g_i \cdot f_{i,y} \cdot Pgcap_{i,y}$$
$$+ \sum_{t \in \mathcal{T}} \sum_{i \in \mathcal{I}} v_{i,y} \cdot Pg_{i,y,t} + \sum_{j \in \mathcal{J}} ASC_{j,y} \quad (2)$$

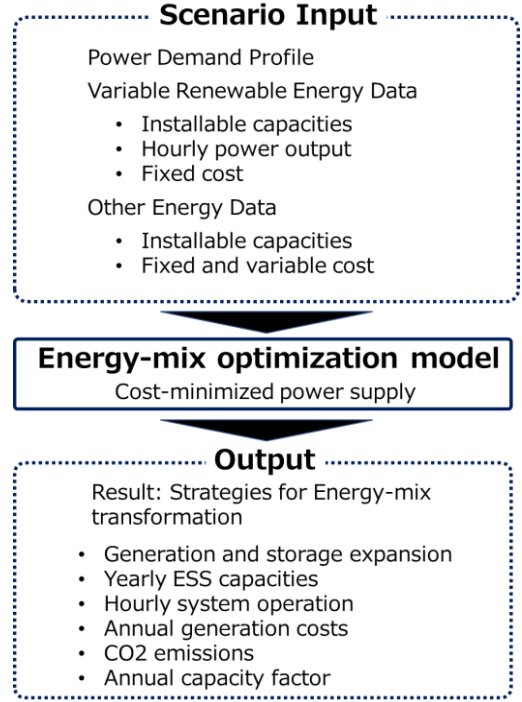

**Fig. 1.** Inputs and outputs of energy-mix optimization model.

$$ASC_{j,y} = gsp_j \cdot fsp_{j,y} \cdot Pscap_{j,y}$$
$$+ gse_j \cdot fse_{j,y} \cdot Escap_{j,y} \quad (3)$$

$$\sum_{i \in \mathcal{I}} Pg_{i,y,t} + \sum_{j \in \mathcal{J}} (Psout_{j,y,t} - Psin_{j,y,t}) = load_{y,t} \quad (4)$$

$$\sum_{i \in \mathcal{I}} Ru_{i,y,t} + \sum_{j \in \mathcal{J}} (Rscd_{j,y,t} + Rsdu_{j,y,t}) \geq rc^+_{y,t} \quad (5)$$

$$rc^+_{y,t} = rcl^+ \cdot load_{y,t} + rcp^- \cdot Pg_{7,y,t}$$
$$+ rcw^- \cdot Pg_{8,y,t} \quad (6)$$

$$\sum_{i \in \mathcal{I}} Rscd_{i,y,t} + \sum_{j \in \mathcal{J}} (Rscu_{j,y,t} + Rsdd_{j,y,t}) \geq rc^-_{y,t} \quad (7)$$

$$rc^-_{y,t} = rcl^- \cdot load_{y,t} + rcp^+ \cdot Pg_{7,y,t}$$
$$+ rcw^+ \cdot Pg_{8,y,t} \quad (8)$$

$$Pg_{i,y,t} + Ru_{i,y,t} \leq Pgcap_{i,y} \quad (9)$$

$$Pg_{i,y,t} - Rd_{i,y,t} \geq 0.0 \quad (10)$$

$$Pg_{i,y,t} = pvg_t \cdot Pg_{7,y} - Pcpv_{y,t} \quad (11)$$

$$Pg_{i,y,t} = wpg_t \cdot Pg_{8,y} - Pcwp_{y,t} \quad (12)$$

$$Ru_{i,y,t} \leq r_i^{+,upper} \cdot Pgcap_{i,y} \quad (13)$$

$$Rd_{i,y,t} \leq r_i^{-,upper} \cdot Pgcap_{i,y} \quad (14)$$

$$Ru_{i,y,t} \leq Pg_{i,y,t} \quad (15)$$

$$Rd_{i,y,t} \leq Pg_{i,y,t} \quad (16)$$

$$Psin_{j,y,t} + Psout_{j,y,t} + Rscu_{j,y,t} + Rsdu_{j,y,t}$$
$$\leq Pscap_{j,y} \quad (17)$$

$$Psin_{j,y,t} - Rscd_{j,y,t} \geq 0.0 \quad (18)$$

$$Psout_{j,y,t} - Rsdd_{j,y,t} \geq 0.0 \quad (19)$$

$$Rscu_{j,y,t} + Rsdu_{j,y,t} \leq rs_j^{+,upper} \cdot Pscap_{j,y} \quad (20)$$

$$Rscd_{j,y,t} + Rsdd_{j,y,t} \leq rs_j^{-,upper} \cdot Pscap_{j,y} \quad (21)$$

$$Es_{j,y,t} \leq Escap_{j,y} \tag{22}$$

$$Es_{j,y,t} \geq 0.2 \cdot Escap_{j,y} \tag{23}$$

$$Es_{j,y,t+1} = \mu s_j \cdot Psin_{j,y,t} - \frac{1}{\mu s_j} \cdot Psout_{j,y,t} + Es_{j,y,t} \tag{24}$$

$$Es_{j,yNum-1,tNum-1} = 0.5 \cdot Escap_{j,yNum-1} \tag{25}$$

$$Es_{j,y,0} = 0.5 \cdot Escap_{j,y} \tag{26}$$

$$Es_{j,y,tNum} = 0.5 \cdot Escap_{j,y} \tag{27}$$

$$Pgcap_{i,y} \leq Pgcap_i^{upper} \tag{28}$$

$$Pscap_{j,y} \leq Pscap_j^{upper} \tag{29}$$

$$Escap_{j,y} \leq Escap_j^{upper} \tag{30}$$

$$Pgcap_{i,y} = \sum_{\tau=0}^{y} Pginst_{i,\tau} + epc_{i,y} \tag{31}$$

$$Pscap_{j,y} = \sum_{\tau=0}^{y} Psinst_{j,\tau} + epsc_{j,y} \tag{32}$$

$$Escap_{j,y} = \sum_{\tau=0}^{y} Esinst_{j,\tau} + eesc_{j,y} \tag{33}$$

$$\sum_{i \in \mathcal{I} \setminus \{7,8\}} Pgcap_{i,y} + \sum_{j \in \mathcal{J}} Pscap_{j,y} \geq (1+\delta) \cdot load_{y,t} \tag{34}$$

$$Pg_{i,y,t+1} \leq Pg_{i,y,t} + d_i^+ \cdot Pgcap_{i,y} \tag{35}$$

$$Pg_{i,y,t+1} \geq Pg_{i,y,t} - d_i^- \cdot Pgcap_{i,y} \tag{36}$$

$$\frac{\sum_{t \in \mathcal{T}} Pg_{i,y,t}}{Pgcap_{i,y} \cdot tNum} \geq cf_i^{min} \tag{37}$$

$$\frac{\sum_{t \in \mathcal{T}} Pgout_{i,y,t}}{Pgcap_{i,y} \cdot tNum} \leq cf_i^{max} \tag{38}$$

, where $i \in \{$ 1: nuclear power, 2: coal-fired power, 3: LNG-fired power, 4: oil-fired power, 5: hydroelectric power, 6: geothermal power, 7: solar power, 8: onshore wind power$\}$

Equation (1) is the objective function. The total cost for the entire period is minimized as the objective function. Equation (2) is a constraint on the annual cost. It sums the annual fixed cost for the generation facility, the annual fuel cost, and the annual fixed cost for the energy storage facility. Equation (3) is a constraint on the one-year fixed cost of the energy storage facility. Equation (4) is the supply-demand balance constraint. Equations (5) and (7) are inequality constraints on the amount of regulation reserve capacity provided by generators. Specifically, they are constraints that ensure more capacity than the required regulation reserve capacity shown in equations (6) and (8). Equation (9) is the generation capacity constraint. Equation (10) is the constraint on the capacity of the regulation reserves in the downward direction. Equations (11) and (12) are constraints on the output of PV and WP. The actual power output minus the amount of curtailment is the respective power output. Equations (13)-(16) are constraints on the upper limit of the capacity of regulation reserves provided relative to the rated capacity. Equation (17) is the constraint of the storage system's recharge/discharge power, power capacity, and regulation reserve capacity. In this paper, we consider many energy storage facilities together. For this reason, constraints on the total capacity of charging and discharging power and the amount of regulating power provided are provided. Equations (18) and (19) are the constraints on the relationship between the charging/discharging power and the downward regulation reserve. Equations (20) and (21) represent the upper limit of the regulation reserve capacity that can be provided relative to the power capacity of the storage system. Equations (22) and (23) are the upper and lower constraints on the amount of stored energy. Equation (24) is the constraint for updating the amount of stored energy between periods. Equations (25)-(27) are equality constraints on the amount of stored energy boundary conditions. In this paper, it is assumed that the amount of stored energy is 50% of capacity at the beginning and end of a year. Equations (28)-(30) are upper bound constraints on the amount of capacity. Equations (31)-(33) are equation constraints on the installed capacity for each year. The capacity in each year is the capacity of the existing facilities plus the newly added capacity. Equation (34) is a constraint on operation reserve capacity. Equations (35) and (36) are upper and lower constraints on the change in generation output. Equation (37) and (38) is the upper and lower constraints on the capacity factor of each generating facility.

**Table 1.** The nomenclatures are used in the energy mix optimization model.

| [Set] | | | [Constants] | |
|---|---|---|---|---|
| | $\mathcal{I}$ | A set of generation types ($\mathcal{I} \in \{1, \ldots, iNum\}$). | $r$ | Discount rate. |
| | $\mathcal{J}$ | A set of storage types ($\mathcal{J} \in \{1, \ldots, jNum\}$). | $g$ | Annual expense rate. |
| | $\mathcal{Y}$ | A set of years ($\mathcal{Y} \in \{1, \ldots, yNum\}$). | $f$ | Fixed cost [JPY/MW]. |
| | $\mathcal{T}$ | A set of time ($\mathcal{T} \in \{1, \ldots, tNum\}$). | $v$ | Variable cost [JPY/MWh]. |
| [Index] | | | $gsp$ | Annual expense rate of power capacity of ESS. |
| | $i$ | Generation type. | $gse$ | Annual expense rate of energy capacity of ESS. |
| | $j$ | Storage type. | $fsp$ | Fixed cost of power capacity of ESS [JPY/MW]. |
| | $y$ | Year. | $fse$ | Fixed cost of energy capacity of ESS [JPY/MWh]. |
| | $t$ | Timeslot in a year. | $load_{y,t}$ | Electrical load [MW]. |
| [Decision variables] All decision variables are non-negative. | | | $pvg_t$ | Generation output of PV [PU]. |
| | $AC$ | Annual cost [JPY]. | $wpg_t$ | Generation output of WP [PU]. |
| | $Pgcap$ | Capacity of generator [MW]. | $rc^+$ | Required capacity of upward regulation reserves [MW]. |
| | $Pg$ | Generation power output [MW]. | $rc^-$ | Required capacity of downward regulation reserves [MW]. |
| | $Pginst$ | Installed capacity of generator [MW]. | $rcl$ | Fluctuation rate of electrical load. |
| | $Psinst$ | Installed power capacity of ESS [MW]. | $rcp^+$ | Upward fluctuation rate of PV. |
| | $Esinst$ | Installed energy capacity of ESS [MWh]. | $rcp^-$ | Downward fluctuation rate of PV. |
| | $ASC$ | Annual storage cost [JPY] | $rcw^+$ | Upward fluctuation rate of WP. |
| | $Pscap$ | Power capacity of ESS [MW] | $rcw^-$ | Downward fluctuation rate of WP. |
| | $Escap$ | Energy capacity of ESS [MWh] | $r^{+,upper}$ | Upper rate of upward regulation reserve. |
| | $Psin$ | Charge power of ESS [MW]. | $r^{-,upper}$ | Upper rate of downward regulation reserve. |
| | $Psout$ | Discharge power of ESS [MW]. | $rs^{+,upper}$ | Upper rate of upward regulation reserve of ESS. |
| | $ES$ | Amount of energy in ESS [MWh] | $rs^{-,upper}$ | Upper rate of downward regulation reserve of ESS. |
| | $Ru$ | Upward regulation reserve capacity of generator [MW]. | $epc$ | Installed capacity of existing generator [MW]. |
| | $Rd$ | Downward regulation reserve capacity of generator [MW]. | $epsc$ | Installed power capacity of existing ESS [MW]. |
| | $Rscu$ | Upward regulation reserve capacity of charging ESS [MW]. | $eesc$ | Installed energy capacity of existing ESS [MW]. |
| | $Rscd$ | Downward regulation reserve capacity of charging ESS [MW]. | $cf^{min}$ | Minimum capacity factor. |
| | $Rsdu$ | Upward regulation reserve capacity of discharging ESS [MW]. | $cf^{max}$ | Maximum capacity factor. |
| | $Rsdd$ | Downward regulation reserve capacity of discharging ESS [MW]. | $\delta$ | Reserve rate. |
| | $Pcpv$ | Curtailment of PV output [MW]. | $d^+$ | Maximum increase rate of power output. |
| | $Pcwp$ | Curtailment of WP output [MW]. | $d^-$ | Maximum decrease rate of power output. |
| | | | $pgcap^{upper}$ | Upper limitation of generator capacity [MW]. |
| | | | $pscap^{upper}$ | Upper limitation of power capacity of ESS [MW]. |
| | | | $escap^{upper}$ | Upper limitation of energy capacity of ESS [MW]. |

## 4. Case Study

In this paper, the energy mix for the Japanese power system was optimized to ascertain the effect of extending the model for the provision of regulation reserve capacity. This case study takes one year as the period of consideration and finds the energy mix that minimizes the total cost for one year.

## 4.1 Input Data

This section describes the data input to the proposed model. The demand data is the sum of hourly electricity demand data for 2019, which is included in the supply-demand data by service area published by the ten general transmission and distribution utilities in Japan.

Assumptions regarding the cost, generator capacity, and installation potential for each type of power generation are shown in Table 2. Data compiled by the Power Generation Cost Verification Working Group [7] was used for fixed costs. Fuel costs, as variable costs, were calculated based on the fuel prices listed in the IEEJ Outlook 2020 [8]. Existing generation capacity was calculated by summing the generation capacity of each power plant disclosed by each company. The minimum and maximum facility utilization rates were set based on the values used by Kawakami et al. [9]. The installable capacity of renewable energies (hydroelectric, geothermal, photovoltaic, and wind power) was calculated using the values published by the Ministry of the Environment [10] as introduction potentials.

**Table 2.** Assumed parameters in a case study.

|  | Nuclear | Coal | LNG | Oil | Hydro | Geothermal | PV | WP |
|---|---|---|---|---|---|---|---|---|
| Fixed cost [$10^4$JPY/kW] [7] | 40.0 | 24.4 | 16.1 | 20.0 | 64.0 | 79.0 | 29.4 | 34.7 |
| Existing generator capacity [MW] | 39,561 | 27,708 | 67,251 | 27,858 | 36,065 | 0.134 | 53,269 | 4,043 |
| Minimum capacity factor [%] [9] | 100 | 30 | 30 | 30 | 11.3 | 0 | - | - |
| Maximum capacity factor [%] [9] | 100 | 85 | 90 | 90 | 55 | 70 | - | - |
| Maximum ramping [%] | 0 | 20 | 20 | 20 | 100 | - | - | - |
| Maximum ramping [%] | 0 | 20 | 20 | 20 | 100 | - | - | - |

The PV and WP output profiles were used based on the PV and WP output recorded in the supply and demand data. They were calculated per unit (pu) values per MW using the installed capacity data as of December 2019 published on the feed-in tariff information website [11]. The rates of change in electricity demand, PV output, and WP output were set at 3%, 10%, and 8%, respectively.

### 4.2 Compared Cases

The model proposed in this paper is unique in that it adds information on the regulation capacity compared to previous studies. In the numerical experiments in this paper, the following two cases were set up to confirm the changes caused by the addition of information on regulation reserve capacity.

- Case 1: Calculate without including constraints on regulation reserve capacity.
- Case 2: Include constraints on regulation reserve capacity (proposed model).

### 4.3 Calculation Results

First, Table 3 shows the optimal installed capacities calculated for the two cases. From Table 3, the WP capacity and the pumped storage energy capacity were increased.

Next, the annual output curtailment is shown in Fig. 2, which shows an increase in PV output curtailment and a decrease in WP output curtailment. This is because the output is suppressed in Case 2 due to the lack of regulation reserve capacity during the hours when PV output is high. In this paper, the output fluctuation rates of PV and WP are set to 10% and 8%, respectively. Therefore, PV was prioritized for output curtailment, and WP output curtailment was reduced.

**Table 3.** Optimal capacity in each case.

|  | Case 1 | Case 2 (Proposed model) |
|---|---|---|
| Nuclear [MW] | 39,561 | 39,561 |
| Coal [MW] | 27,708 | 27,708 |
| LNG [MW] | 67,251 | 67,251 |
| Oil [MW] | 27,858 | 27,858 |
| Hydro [MW] | 36,065 | 36,065 |
| Geothermal [MW] | 10,375 | 10,375 |
| PV [MW] | 53,269 | 53,269 |
| WP [MW] | 76,288 | 76,651 |
| Pump (Power) [MW] | 26,000 | 26,000 |
| Pump (Energy) [MWh] | 130,000 | 177,518 |
| BESS (Power) [MW] | 1,600 | 1,600 |
| BESS (Energy) [MWh] | 9,600 | 9,600 |

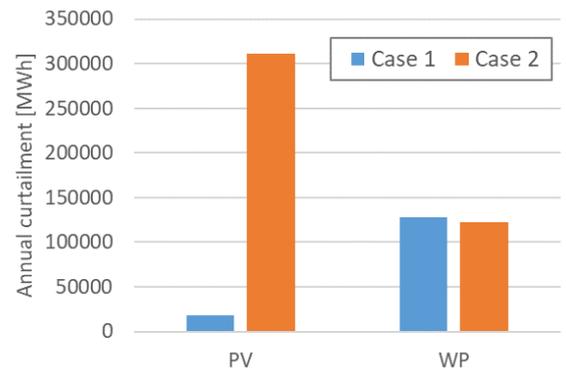

**Fig. 1.** Annual curtailment of PV and WP.

As shown in Table 3, the increase in pumped storage energy capacity and WP capacity can be attributed to the fact that the PV output curtailment was covered by WP and pumped storage.

However, the results of numerical experiments conducted by the author with similar input data in the literature [12] showed a minor change in PV and WP output suppression when regulation reserve capacity was included in the model. This is because the model proposed in this paper lumps all areas together. In the existing power system, each area is connected by interconnection lines, and each area has a different power supply configuration. Therefore, the regional bias of renewable energy dramatically affects the amount of output curtailment. The results of this paper confirm the necessity of constraints on cross-regional interconnection lines.

## 5. Conclusion

In this paper, a long-term energy mix optimization model is proposed. A case study for Japan confirms that generators' operation and optimal capacity changed with the addition of the constraint of regulation reserve capacity.

In Japan, interconnection lines connect each area with a different energy mix and renewable energy installable capacity. Literature [12] proposes an efficient method for solving the UC problem for multiple areas connected by interconnection lines. It is thought that the effect of the model proposed in this paper, including the regulation reserve capacity, can be achieved by introducing interconnection lines as a model.

This paper compares results from a single year to confirm the effect of the model extension but does not calculate or compare results from multiple years. In the future task, it will be necessary to conduct multi-year case studies and study facilities' transitions to reach the $CO_2$ emission reduction target.